# THERMAP: a mid-infrared spectro-imager for space missions to small bodies in the inner solar system

O. Groussin[1], J. Licandro[2], J. Helbert[3], J.-L. Reynaud[1], P. Levacher[1], M. Reyes García-Talavera[2], V. Alí-Lagoa[2], P.-E. Blanc[1], E. Brageot[1], B. Davidsson[4], M. Delbó[5], M. Deleuze[6], A. Delsanti[1], J. J. Diaz Garcia[2], K. Dohlen[1], D. Ferrand[1], S. F. Green[7], L. Jorda[1], E. Joven Álvarez[2], J. Knollenberg[3], E. Kührt[3], P. Lamy[1], E. Lellouch[8], J. Le Merrer[1], B. Marty[9], G. Mas[6], C. Rossin[1], B. Rozitis[7], J. Sunshine[10], P. Vernazza[1] and S. Vives[1]

[1] Aix Marseille Université, CNRS, LAM (Laboratoire d'Astrophysique de Marseille) UMR 7326, 13388, Marseille, France
[2] Instituto de Astrofísica de Canarias, c/Via Lactea s/n, 38200 La Laguna, Tenerife, Spain
[3] Deutsches Zentrum für Luft-und Raumfahrt (DLR), Berlin, Germany
[4] Department of Physics and Astronomy, Uppsala University, Uppsala, Sweden
[5] Université de Nice Antipolis, CNRS, Observatoire de la Côte d'Azur, Nice, France
[6] Centre National d'Etudes Spatiales (CNES), Toulouse, France
[7] Planetary and Space Sciences, Department of Physical Sciences, The Open University, Milton Keynes, UK
[8] Laboratoire d'Études Spatiales et d'Instrumentation en Astrophysique (LESIA), Observatoire de Paris, CNRS, UPMC, Université Paris Diderot, Meudon, France
[9] Centre de Recherches Pétrographiques et Géochimiques, Vandoeuvre-lès-Nancy, France
[10] University of Maryland, Department of Astronomy, College Park, MD, USA

## Abstract
We present THERMAP, a mid-infrared spectro-imager for space missions to small bodies in the inner solar system, developed in the framework of the MarcoPolo-R asteroid sample return mission. THERMAP is very well suited to characterize the surface thermal environment of a NEO and to map its surface composition. The instrument has two channels, one for imaging and one for spectroscopy: it is both a thermal camera with full 2D imaging capabilities and a slit spectrometer. THERMAP takes advantage of the recent technological developments of uncooled microbolometer arrays, sensitive in the mid-infrared spectral range. THERMAP can acquire thermal images (8 – 18 µm) of the surface and perform absolute temperature measurements with a precision better than 3.5 K above 200 K. THERMAP can acquire mid-infrared spectra (8 – 16 µm) of the surface with a spectral resolution Δλ of 0.3 µm. For surface temperatures above 350 K, spectra have a signal-to-noise ratio >60 in the spectral range 9 – 13 µm where most emission features occur.

## Keywords
Mid-infrared, spectro-imager, microbolometer, MarcoPolo-R, ESA's Cosmic Vision, sample return

## Corresponding author
Olivier Groussin
Laboratoire d'Astrophysique de Marseille
38 rue Frédéric-Joliot Curie, 13388 Marseille Cedex 13, France
Tel: (+33) 491 056 972 / Email: olivier.groussin@lam.fr

## Acknowledgements
The contribution from France to the THERMAP instrument was funded by the Centre National d'Etudes Spatiales (CNES). The contribution from Spain to the THERMAP instrument was funded by the Spanish "Ministerio de Economía y Competitividad" projects AYA2011-29489-C03-02 and AYA2012-39115-C03-03. M. Delbo acknowledges support from the French "Agence Nationale de la Recherche" (ANR-SHOCKS).

## 1   Introduction

Near-Earth Objects (NEOs) are solar system bodies with a perihelion distance q<1.3 AU. As with other small bodies in the solar system like main belt asteroids, comets or trans-Neptunian objects, NEOs are residuals of the early solar system and provide important constraints on its formation and evolution. The interest in NEOs also resides in the fact that they come close to the Earth's orbit and are thus a potential threat to Earth.

More than 12 500 Near-Earth Objects (NEOs) have currently been detected but only 10% of them have been observed in the mid-infrared (5-25 µm) from ground or space-based telescopes, mainly with WISE through the NEOWISE program (Mainzer et al. 2011) and with Spitzer through the ExploreNEOs program (Trilling et al. 2010). However, none have been spatially resolved or observed in situ in the mid-infrared window. This wavelength range, which probes thermal emission, remains an unexplored field for space missions to NEOs. The THERMAP instrument, a mid-infrared spectro-imager for space missions to small bodies in the inner solar system, has been specifically designed to fill this gap.

We developed THERMAP to fulfil three main scientific objectives: 1) characterize the surface thermal environment of a NEO, 2) map the surface composition of a NEO and 3) put the sample within its context, in the framework of an asteroid sample return mission. THERMAP takes advantage of the recent technological developments of microbolometers. These uncooled detectors, sensitive to radiation in the mid-infrared, are now available in large arrays (up to 1024x768 pixels), small pixel size (17 µm) and high sensitivity (Noise Equivalent Temperature Difference or NETD<60 mK at 300K with a focal ratio of 1). This makes microbolometers very attractive. In particular, we wanted to take advantage of these large 2D arrays to build an instrument with imaging capabilities, i.e. a thermal camera similar and complementary to classical visible cameras, *and* with spectroscopic capabilities.

The THERMAP instrument was selected in February 2013, following an announcement of opportunity for the scientific payload of the ESA MarcoPolo-R mission. MarcoPolo-R is an asteroid sample return mission (Barucci et al. 2012), down selected after the assessment phase of the 4th medium class mission (M4). Although THERMAP will not fly on MarcoPolo-R, this instrument concept is promising and could be proposed for future space missions towards airless bodies of the inner solar system. Mid-infrared spectro-imagers are indeed scientifically very attractive, as demonstrated by studies of the OSIRIS-REx Thermal Emission Spectrometer (OTES) onboard the OSIRIS-REx mission, and inclusion on strawman payloads for future proposed small bodies missions such as ESA's Asteroid Intercept Mission (AIM) and Phobos Sample Return.

In this paper we present the results of the assessment phase of the THERMAP instrument, performed in 2013 by the Laboratoire d'Astrophysique de Marseille (LAM, France – PI: O. Groussin), the Instituto de Astrofísica de Canarias (IAC, Spain – Lead scientist: J. Licandro), and the Deutsches Zentrum für Luft-und Raumfahrt (DLR, Germany – Lead scientist: J. Helbert). In Sect. 2 we give an overview of the scientific objectives of THERMAP. In Sect. 3 we describe the instrument. Conclusions and perspectives are given in Sect. 4.

## 2   Scientific objectives of the THERMAP instrument

The scientific objectives of the THERMAP instrument are to *characterize the surface thermal environment of a NEO* and to *map the surface composition of a NEO*. In particular, the aim of THERMAP is to answer the following questions: what is the surface temperature and its spatial and temporal variability? What are the bulk thermal properties of the surface, such as thermal inertia? What is the surface roughness and does it vary spatially? What is the degree of thermal stress weathering? What is the surface mineralogical composition? What is the effect of space weathering on surface composition? To what extent can we constrain the Yarkovsky and YORP effects for NEOs?

In the framework of an asteroid sample return mission, an additional scientific objective is to *put the sample within its context*, in terms of thermal environment and surface composition of the sampling site and surroundings. For operations, THERMAP will also help to select the landing site. As an example, the thermal inertia is a good proxy for the presence or absence (i.e., bare rocks) of regolith on the surface, which is fundamental for the sampling mechanism that can handle dust and gravel, but not a consolidated material of 10's of cm or more.

The scientific objectives of THERMAP are summarized in the science traceability matrix (Table 1).

**Table 1** – The THERMAP science traceability matrix.

| Science goals and questions | | Measurement objectives | Observing mode | | Spatial resolution | Science results |
| --- | --- | --- | --- | --- | --- | --- |
| | | | *Imaging* | *Spectro* | | |
| **Characterize the surface thermal environment of a NEA** | What is the surface temperature and degree of thermal stress weathering? | Map the surface temperature of the asteroid, several times per rotation, and of the sampling site, pre and post sampling | ✔ | | 10 m to 5 m | - Diurnal and seasonal thermal cycles<br>- Influence of shape and topography<br>- Degree of thermal stress weathering |
| | What are the bulk thermal properties of the surface? | Map the surface temperature of the asteroid, several times per rotation | ✔ | | 10 m to 5 m | - Thermal inertia<br>- Surface roughness<br>- Yarkovsky and YORP effects |
| **Map the surface composition of a NEA** | What is the surface mineralogical composition? | Map the surface composition of the asteroid | | ✔ | 10 m to 5 m | - Mineralogical composition of the surface and surface heterogeneities |
| | What is the effect of space weathering on surface composition? | Map the surface composition of the sampling site, pre and post sampling | | ✔ | 0.25 m | - Space weathering effects |
| **Select the sampling site and put the sample within its context** | What is the surface thermal environment of the sampling site? | Map the surface temperature of the sampling site | ✔ | | 0.25 m | - Thermal environment of the sampling site compared to the rest of the surface |
| | What is the surface composition of the sampling site? | Map the surface composition of the sampling site | | ✔ | 0.25 m | - Mineralogical composition of the sampling site compared to the rest of the surface |

## 2.1 Characterize the surface thermal environment of a NEO

### 2.1.1 Surface temperature and thermal stress weathering

The Sun is the primary source of energy for NEOs and drives the surface energy balance, which controls the temperature. The temperature of a NEO varies across the surface, spatially and temporally. For a NEO at 1 AU from the Sun, the surface experiences large and fast temperature fluctuations, passing every few hours from >400 K close to the sub-solar point to <150 K on the night side. This extreme thermal environment leads to thermal stress weathering of the surface, which might change the primordial surface structure of the NEO (Delbo et al. 2014).

#### 2.1.1.1 Temperature on the surface

Temperature varies spatially across the surface, depending mainly on global asteroid shape, local topography and bulk thermal properties (Fig. 1). The surface temperature also changes with time, on short time scales for the diurnal variations (e.g., hours) and on long time scales for the seasonal variations (e.g., months). So, the temperature on the surface provides important information to: i) characterize the diurnal and seasonal thermal cycles and how temperatures varies across the surface, with the minimum, maximum and average temperature and the rate of temperature change with time, ii) study the influence of global shape and local topography on the diurnal and seasonal cycles, iii) constrain the surface energy budget and the physical processes occurring on the surface (mainly solar absorption, thermal emission and heat conduction) and iv) characterize the bulk thermal properties of the surface.

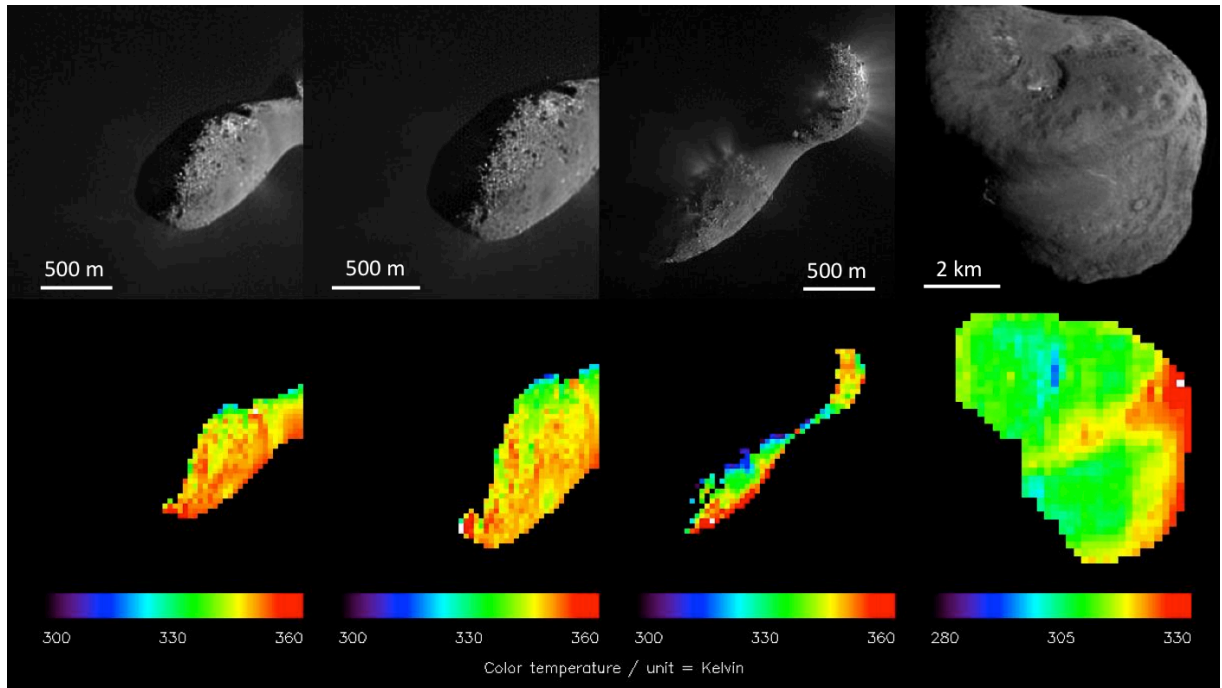

**Fig. 1** – Images (top panels) and surface color temperature (bottom panels) of the nucleus of comet 103P/Hartley 2 (3 left images) and 9P/Tempel 1 (right image), acquired by the Deep Impact spacecraft (Groussin et al. 2013).

#### 2.1.1.2 Temperature anomalies

The surface temperature distribution is not uniform and the identification of cold or hot spots, i.e. spots that are significantly colder or hotter than their surroundings, can be used as a proxy to track a difference in solar insulation conditions, composition, albedo or thermal properties. Cold and hot spots are scientifically interesting since they reveal regions with a significantly different thermal environment than the average NEO, which could be, for the extremes, in permanent shadow or constant illumination. They can also reveal terrains with different degrees of compaction, as in the case of the Moon observed by Price et al. (2003) with the Midcourse Space Experiment (MSX). Cold or hot spots can be detected at small scales (small craters and depressions, grooves, holes, walls, pits, boulders, …) and large scales (large craters and depressions, mountains, polar caps, …).

#### 2.1.1.3 Thermal stress weathering

The surface experiences large diurnal and seasonal thermal cycles, with temperature changes up to several hundred Kelvin every rotation. These cycles, repeated millions of times since the formation of the NEO, represents a strong thermal stress for the surface, which may have been significantly altered. The question naturally arises as to what extent thermal processing changes the primordial record of the NEO surface. Delbo et al. (2014) demonstrated with laboratory experiments that thermal stress is the dominant process governing regolith generation on small asteroids. On a real NEO, the degree of thermal stress weathering and its efficiency is however unknown and requires the observation of the same region, before and after refreshing the surface. This would be possible for the sampling site of a sample return mission, which will be observed before and after sample collection. This however requires a sampling mechanism able to "brush" the upper surface layers, exposing fresh materials. Assuming that thermal stress weathering is primarily controlled by the diurnal temperature variations, a layer of several centimetres should be removed, which corresponds to the typical diurnal thermal skin depth of asteroids (Harris and Lagerros 2002).

#### 2.1.1.4 The Yarkovsky and YORP effects

Another important scientific objective is to quantify the Yarkovsky and YORP effects (named after Yarkovsky, O'Keefe, Radzievskii and Paddack) for NEOs (Rubincam 2000; Morbidelli and Vokrouhlicky, 2003). These two effects result from the non-gravitational force due to the emission of thermal photons, which carries momentum. For a body with non-zero thermal inertia, the thermal emission is anisotropic and has a net component along the orbital velocity vector and along the angular momentum of the body. This force (Yarkovsky) and couple (YORP) change the orbital and rotational parameters of the asteroid, respectively.

Yarkovsky and YORP effects are important to constrain the long-term evolution of asteroids. This is particularly relevant in the case of potentially hazardous asteroids, the prediction of their orbits and their possible deviation to avoid collision with the Earth. The Yarkovsky and YORP effects are controlled by the surface temperature distribution, which can be acquired by space missions with sufficient spatial resolution and accuracy to properly constrain and disentangle these two effects (Rozitis and Green 2013).

### 2.1.2 Bulk thermal properties of the surface

The temperature on the surface and its diurnal variations depend on several physical (albedo, rotation axis, rotation period, etc.) and thermal quantities. For the thermal quantities, measurable by a mid-infrared instrument, this mainly includes thermal inertia and surface roughness at the sub-pixel scale. Since the surface temperature is a proxy for both of them, thermal inertia and surface roughness are often called thermal properties, even if strictly speaking surface roughness is a topographic property.

#### 2.1.2.1 Thermal inertia

Thermal inertia drives the ability of a material to adapt its temperature to a change in energy input, i.e., insolation for asteroids. Thermal inertia $\Gamma$ is defined by $\Gamma = \sqrt{\kappa \rho C}$, where $\kappa$ is the heat conductivity, $\rho$ the density and $C$ the heat capacity. For a small thermal inertia (typically $\Gamma < 50$ W K$^{-1}$ m$^{-2}$ s$^{-1/2}$), the surface temperature adapts almost instantaneously to a change in insolation, while for a large thermal inertia (typically $\Gamma > 1000$ W K$^{-1}$ m$^{-2}$ s$^{-1/2}$), it takes a longer time for the surface to adjust its temperature. Fig. 2 illustrates this phenomenon and shows how the thermal inertia can be derived from the diurnal thermal cycle, assuming $\kappa$ and $C$ are temperature invariants. It is important to mention that in reality conductivity $\kappa$ depends on temperature, and therefore thermal inertia.

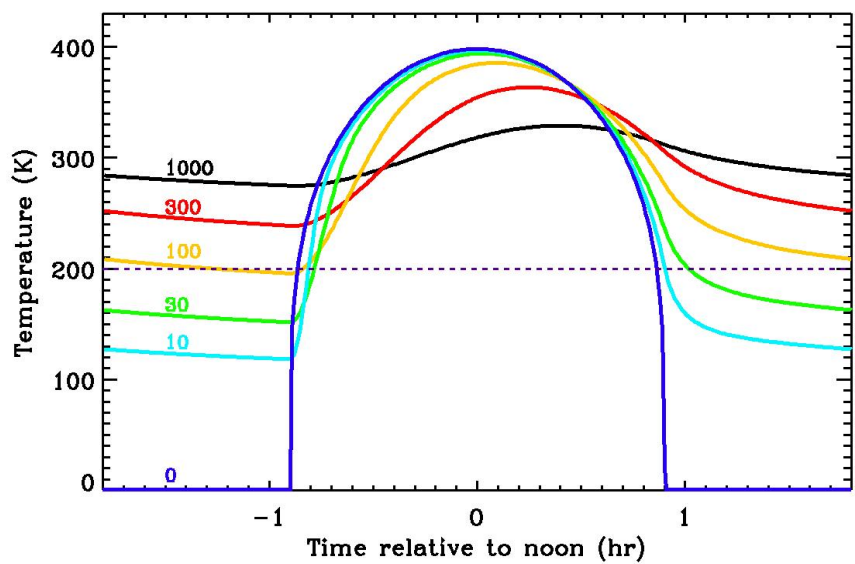

**Fig. 2** – Diurnal thermal cycles on the surface of a NEO at 1 AU from the Sun, for different values of the thermal inertia between 0 and 1000 W K$^{-1}$ m$^{-2}$ s$^{-1/2}$. Temperatures correspond to a surface element on the equator of an asteroid with a rotation period of 3.6 hours and a pole orientation perpendicular to its orbital plane. The surface is assumed to be smooth (no surface roughness). The horizontal dashed line corresponds to the estimated lower limit of 200 K for accurate temperature measurements with THERMAP (Sect. 3.3.1.3.).

Thermal inertia indicates how much solar energy penetrates into the interior and to what depth (the thermal skin depth), which is fundamental to understand to what extent the surface layers are thermally altered. This point is particularly relevant for a sample return mission that will collect surface materials from the first centimetres.

Finally, thermal inertia provides important information on the nature of the surface regolith, and in particular its degree of maturity and thickness (grain size, porosity; Gundlach and Blum 2013). As illustrated in Fig. 3, a surface with a thick, mature, regolith like the Moon usually has a lower thermal inertia than a surface with a shallow/coarser regolith like Eros or even a surface with less regolith like Itokawa.

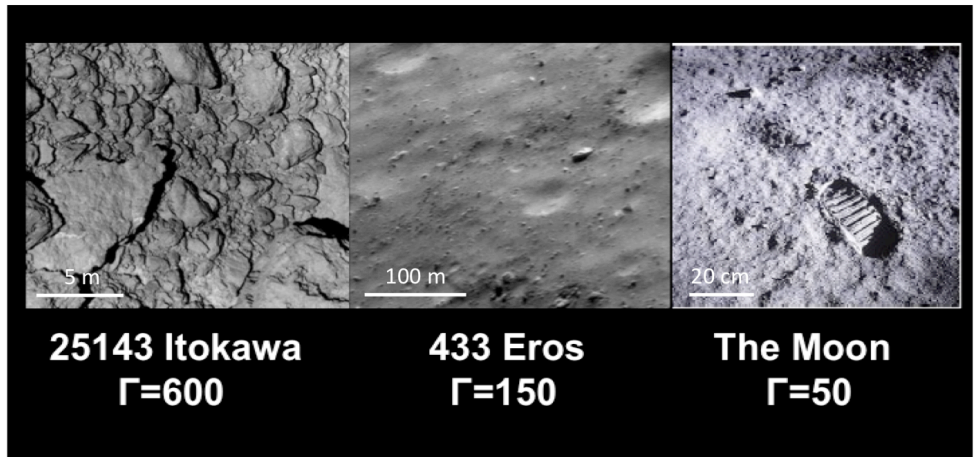

**Fig. 3** – Images of the regolith of several asteroids with different thermal inertia Γ (W K$^{-1}$ m$^{-2}$ s$^{-1/2}$). Scale bars are indicative.

### 2.1.2.2  Surface roughness

The surface of NEOs is not flat but rough. Surface roughness and its associated effects like projected shadows play an important role on the spatial distribution of surface temperatures at the sub-meter scale and more generally at all scales below the spatial resolution of the measurements (Groussin et al. 2007, 2013; Rozitis and Green 2011, 2012; Davidsson et al. 2013), down to the scale at which lateral heat conductivity becomes non-negligible, i.e. the centimetre scale.

Depending on the shape and intensity of the spectral energy distribution (SED), which results from the sub-pixel surface temperature distribution (Fig. 4), it is possible to estimate the degree of roughness. For this scientific objective, in-situ observations are however mandatory since they provide sufficient spatial resolution and allow control of the geometry (incident and emission angles) to remove ambiguities in the directional behaviour of thermal radiation.

As for roughness, thermal inertia also changes the shape of the SED (Müller 2002). Therefore, a specific observational strategy is required to properly estimate surface roughness and in particular to break the degeneracy between thermal inertia and roughness effects (Davidsson et al. 2015).

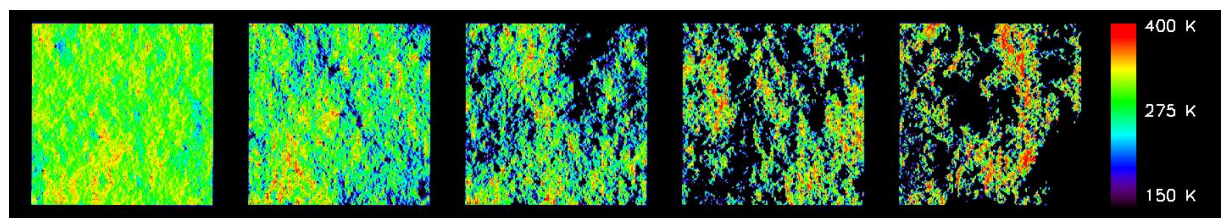

Fig. 4 – Computed temperature distribution on the surface of an artificial NEO, for increasing roughness from left to right (rms slope of 12°, 22°, 36°, 52° and 66°, respectively). Each image shows the temperature of an artificial fractal terrain made of 33 000 facets (Davidsson et al. 2015) and illuminated from the left side with an incidence angle of 60° (nadir observation). For very large roughness, projected shadows become important (more black, non illuminated, facets) and a larger surface area reaches the maximum

temperature (more facets are directly orientated towards the Sun). If each image actually corresponds to one pixel of the THERMAP instrument (e.g., 5 m × 5 m from an altitude of 10 km), they would show different spectral energy distributions in terms of shape and intensity, illustrating the effects of roughness at the sub-pixel scale (Groussin et al. 2013).

## 2.2 Map the surface composition of a NEO

### 2.2.1 Surface mineralogical composition

Mid-infrared spectroscopy can provide compositional information that is not accessible in other spectral domains. Indeed, most major mineral groups found in meteorites (the best analogues of asteroidal surfaces that we have in our laboratory after the grains of Itokawa returned by the Hayabusa mission) and silicate glasses (e.g., maskelynite), which lack useful diagnostic features at visible to near-IR wavelengths (VNIR; 0.4-2.5 µm), do produce diagnostic mid-infrared features. There is therefore a strong interest in exploring the spectral properties of asteroids that are featureless in the VNIR range at those longer wavelengths.

Recent studies (Vernazza et al. 2011, 2012) have demonstrated the paramount importance of the mid-infrared domain to constrain the surface composition of asteroids. Asteroid emission features in the thermal infrared allowed Vernazza et al. (2011) to infer that the spectrum of asteroid Lutetia could be matched with that of enstatite chondrites, thought to originate from the terrestrial region of the protoplanetary disk (≈1 AU) instead of the main asteroid belt. Moreover, mid-IR emission features of Jupiter Trojans and other large (D>200km) asteroids belonging to the spectroscopic C-complex suggest an important surface porosity (>90%) for the first millimetre (Vernazza et al., 2012). It therefore appears that mid-IR emission spectra of asteroids not only carry information about their surface composition but can also help to constrain their surface structure.

The 8-16 µm range is particularly well suited for studying the composition of silicates as it contains the following signatures: (i) residual Reststrahlen features, which occur as emissivity bands (e.g., 10 µm), (ii) overtone/combination tone bands, which occur as emissivity peaks, and (iii) the Christiansen feature, which occurs as a peak in emissivity. Importantly, this spectral domain also allows distinction between the crystalline and/or amorphous nature of silicates, which is of primary importance for constraining the thermal evolution of an asteroid.

Retrieving the composition of a surface from its spectrum remains difficult and mid-IR spectra are not an exception in this domain. Depending on the geometric observational conditions (incidence and emergence angles), while the position of the main emission features does not change to first order, their shape and depth is modified significantly. Laboratory measurements are therefore mandatory to properly interpret spectra and THERMAP will benefit from the work of Maturilli et al. (2008) in building a large mid-IR emissivity database.

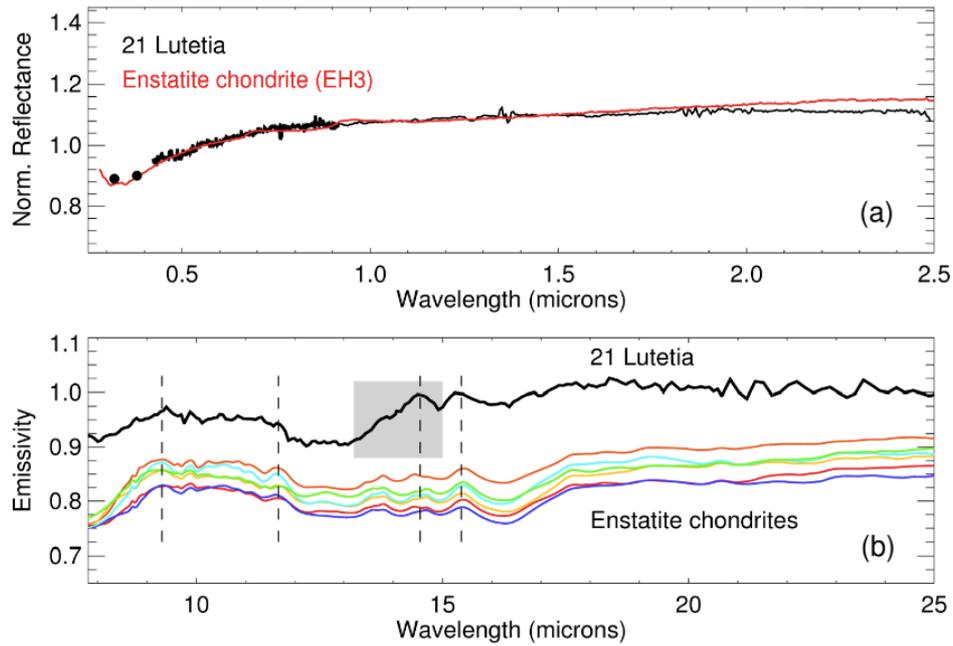

**Fig. 5** – Spectroscopy in the mid-infrared domain revealed the meteoritic analog of the asteroid (21) Lutetia, namely enstatite chondrites (Vernazza et al. 2011). While Lutetia's spectrum appears featureless in the 0.4-2.5 µm range (upper panel), its emissivity spectrum and in particular the 8-16 µm range (lower panel) contains many features that allowed its surface composition to be constrained and identification of its meteoritic analog. The vertical dashed lines correspond to emissivity peaks, present in the asteroid *and* meteorite spectra.

### 2.2.2 Space weathering

Space weathering is the generic term for the processes (irradiation by cosmic and solar wind ions and bombardment by interplanetary dust) that modify the optical properties of surfaces of bodies lacking atmospheres when exposed to the space environment (Fig. 6). Its effect on the interpretation of asteroid spectral data and on the identification of potential linkages between asteroids and meteorites has been an area of considerable discussion and contention for three decades (Chapman 2004). Both laboratory experiments (Strazzulla et al. 2005, Brunetto et al. 2005, 2006a, 2006b; Marchi et al. 2005; Vernazza et al. 2006), ground based observations (Vernazza et al. 2009a) and the Hayabusa mission (Hiroi et al. 2006, Nakamura et al. 2011) have ultimately demonstrated that space weathering processes similar to those acting on the Moon (Pieters et al. 2000) can redden and darken the spectrum of fresh Ordinary Chondrite (OC) or Howardite-Eucrite-Diogenite (HED) like asteroid surfaces, thus bringing to an end a thirty year long conundrum.

The above argument however only applies to the spectral slope and many additional effects of space weathering (e.g., albedo, color) are still unclear (Fig. 6). Recent laboratory experiments simulating the solar wind irradiation have also shown that certain meteorite classes (mesosiderite, enstatite chondrite) remain relatively unaffected by space weathering processes, thus simplifying the search for their parent asteroids (Vernazza et al. 2009b). Yet, for most classes of asteroids and meteorites, space weathering remains poorly understood, mainly because of the scarcity of dedicated experiments (Fulvio et al. 2012).

Imaging and spectroscopy of the surface of an asteroid with high spatial resolution will allow us to characterize the regolith alteration. Visible and thermal images will, in a first step, help us distinguish young regions from older ones (by computing a map of gravitational slopes) and, in a second step, allow us to measure the spatially resolved physical properties (albedo, thermal inertia, roughness) for surfaces of different ages. In turn, these maps will provide a context for the mid-infrared spectral data so that the spectral effects induced by space weathering processes can be fully investigated. Finally, future sample return missions will, as previously done by Hayabusa in the case of the S-type asteroid Itokawa, allow us to quantify space weathering by comparing the sampling site before and after sampling ("fresh" surface). This however requires a sampling mechanism able to "brush" the upper surface layers (cm's), moreover on an area that is large enough to be detected by the THERMAP instrument, i.e. typically 20 cm or more.

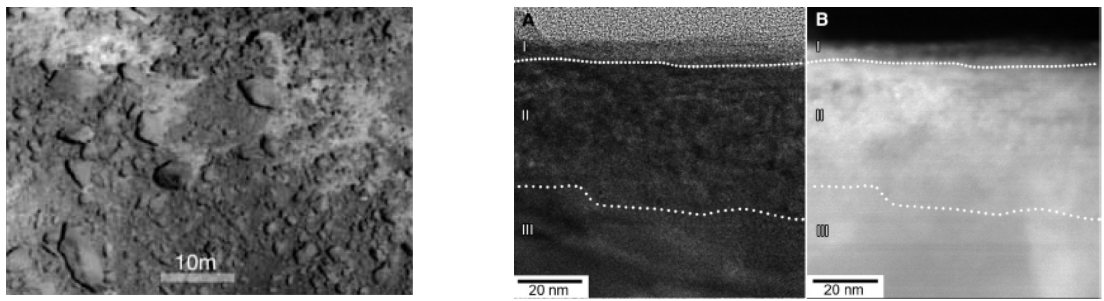

**Fig. 6 –** *Left panel* (Saito et al. 2006): high-resolution image of Itokawa's surface taken from the Hayabusa spacecraft. Space weathering processes are interpreted to be the cause of the albedo (color) variation. *Right panel* (Noguchi et al. 2011): Edge-on BFSTEM and HAADF-STEM images of (A and B) olivine in an Itokawa sample. The rims are divided into three zones based on their texture: zone I, amorphous surface layer containing nanophase iron (npFe, which provides clear evidence of space weathering); zone II, partially amorphized area containing still abundant npFe; zone III, crystalline substrate minerals containing no npFe, consistent with a fresh unweathered interior.

## 2.3    Select the sampling site and put the sample within its context

A key point for the success of a sample return mission is the detailed characterization of the sampling site to put the sample in its global context. It is fundamental to assess to what extent the sampling site is representative of the rest of the asteroid, before any scientific conclusions from the sampling site could be generalized. The characterization of the sampling site includes the determination of its thermal environment (Sect. 2.1) and composition (Sect. 2.2). This characterization will also provide direct information on the temperature distribution and composition heterogeneities at small scales (0.25 m) compared to large scales (5 to 10 m for global characterization), as illustrated in Fig. 7. Finally, the thermal environment of the sampling site is also important for technical reasons, to select a landing site in accordance with the spacecraft capabilities (e.g., thermal constraints).

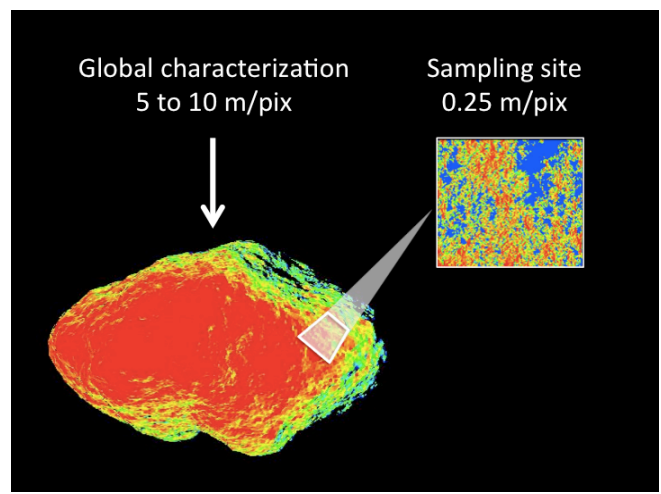

**Fig. 7 –** Sampling site spatial resolution compared to global characterization.

## 3 Description of the THERMAP instrument

### 3.1 Scientific requirements

The design and performance of the THERMAP instrument are driven by the above scientific objectives (Sect. 2), which we converted into three scientific requirements for THERMAP:
   #1. The surface temperature of the complete NEO shall be derived to an accuracy of at least 5 K (goal 1 K) above 200 K. The spatial resolution (= twice the pixel scale) shall be of the order of 10 m at a number of rotational phases from which the thermal inertia can be determined to a precision of better than 10 %.
   #2. The complete surface of the NEO shall be imaged with a spatial resolution of the order of 10 m or better and with a spectral resolution $\lambda/\Delta\lambda$ of the order of 30 or better to determine the wavelength dependent emissivity, and hence identify mineral features in the range 8 – 16 µm (goal 5 – 25 µm).
   #3. In the framework of an asteroid sample return mission, a representative area of the sampling site shall be imaged with a spatial resolution of decimetres and a spectral resolution $\lambda/\Delta\lambda$ of the order of 30 or better to determine the wavelength dependent emissivity, and hence identify mineral features in the range 8 – 16 µm (goal 5 – 25 µm).

### 3.2 Instrument design

Trade-off studies and simulations led to the choice of the following instrument design solution, illustrated in Fig. 8:
- A pointing mirror at the entrance of the instrument, which allows pointing alternatively at the asteroid and three calibration sources (deep space and two black bodies).
- A catadioptric telescope, following the pointing mirror and based on a TMA (Three Mirror Anastigmat) design, which is used by two channels:
    - The imaging channel,
    - The spectroscopic channel, using a classical grating spectrometer, based on an Offner relay design.
- A flip mirror, which allows switching between imaging and spectroscopy.
- Two uncooled microbolometer arrays, one for each channel.
- A command and control unit, including detector Front End Electronics (FEE) and Power Supply Unit (PSU), which interfaces with the spacecraft.

The two instrument apertures, oriented towards the asteroid (for observation) and deep space (for calibration), are protected from stray light entrance by optical baffles.

In imaging mode, the scene is focused on the FPA-I detector. In this configuration, the flip mirror placed in front of the FPA-I detector is removed from the optical beam and placed in a "parking" position by a bi-stable mechanism. In spectroscopic mode, the flip mirror is rotated, so that it intercepts the light beam in front of the FPA-I detector and reflects it onto the entrance slit of the spectrometer, composed of a classical Offner relay and a second detector, mounted on FPA-S. The FPA-S detector is identical to the FPA-I detector.

A stepper motor drives the pointing mirror mechanism. A DC brushless torque motor drives the flip mirror. Command and control functions include pointing unit electronics, calibration electronics, flip mirror electronics, detector unit electronics and power supply unit.

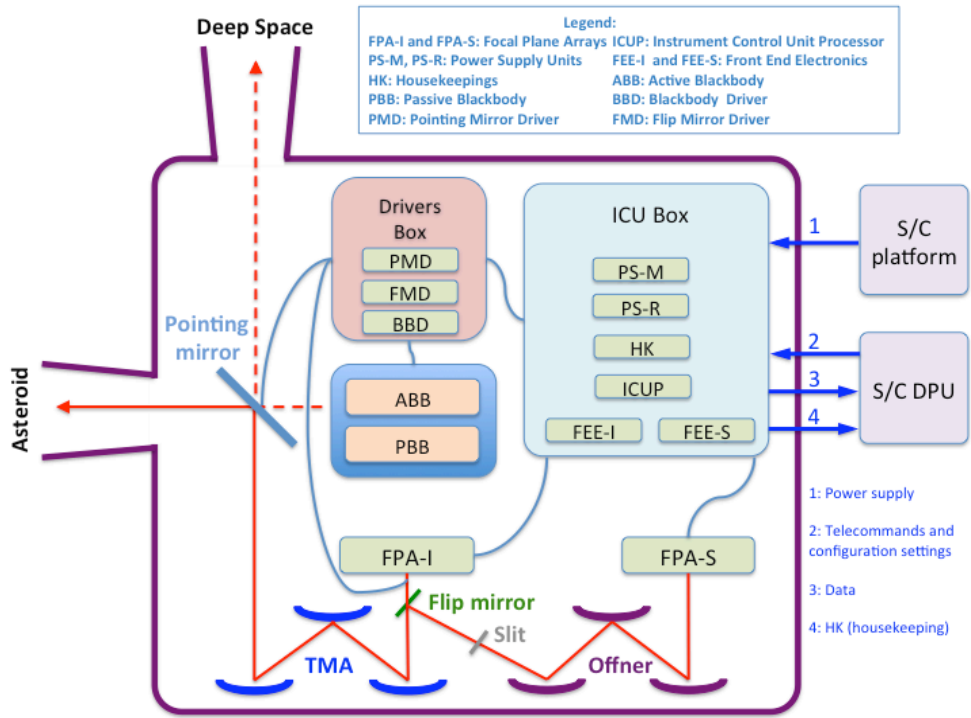

**Fig. 8 –** Block diagram of the THERMAP instrument.

### 3.2.1 Detectors

The heart of the instrument is composed of two large 2D microbolometer arrays, sensitive to mid-IR thermal emission (typically 8 to 18 µm). Our current best candidate is an uncooled microbolometer array with 640x480 pixels and a pixel size of 25 µm. Such detectors are available from the ULIS Company (Grenoble, France). The technology of microbolometer detectors has already been used by several space missions (e.g., Mars Odyssey, LCROSS, Bepi-Colombo) and is thus at a high TRL (Technology Readiness Level). The specific detector used for the THERMAP instrument is also being used by the IAC for the JAXA Extreme Universe Space Observatory (JEM-EUSO) project. IAC has performed extensive tests to characterize this detector (Martin-Hernando et al. 2014), which currently has a TRL of 5, i.e. the detector has been validated in a relevant environment. The microbolometer is optimized to operate at a temperature of 20 – 30°C. More important, it requires an excellent thermal stability of 10 mK rms, controlled by use of a Peltier cell inside it. A detailed characterisation of the ULIS 640x480 detector can be found in Brageot et al. (2014).

### 3.2.2 Optical, mechanical and thermal design

The optical design (Fig. 9 and Table 2) is classical and built around reflective components: TMA miniaturized optics and Offner type spectrometer, which have been extensively studied (e.g., Dohlen et al. 1996) and are now commonly used for infrared spectro-imager instruments. Reflective optics provide a relatively compact achromatic and athermal system, with a very good transmission value, which is crucial for the spectroscopic performance. Classical spectro-imager instruments, either based on direct dispersion (DIS) or Fourier transform (FTS) design, usually have only one channel (for spectroscopy) and require an internal (scanning mirror) or external (pushbroom) scanning device to perform imagery. On the contrary, THERMAP combines a classical imaging channel and a spectrometer in a single instrument, the only mechanism being a single 2 position flip mirror, which allows switching between the two modes. This point is particularly important for a mission to a small asteroid (size of a few hundred meters), for which pushbroom is problematic (the rotational parameters are unknown and there may not be any stable orbits around the asteroid) and for which flexibility for operations is mandatory (i.e., real 2D imagery like a classical camera).

The mechanical design (Fig. 9) is strongly constrained by thermal considerations, leading to the choice of a mechanical structure in aluminium, thermally well connected to the spacecraft panel facing the sky. The electronics box (e-box) is connected close to this interface to allow good thermal conduction from the e-box to the spacecraft, which is used as a radiator. The zone that includes the e-box is called the "warm zone". The optical bench, supporting the whole optics, is thermally isolated from the main structure to ensure a passive low pass filter against temperature variations observed in the structure, and in particular in the warm zone. The zone that includes the optics is called the "cold zone". A mechanical wall, which forms part of the structure, separates the two zones. Its surface has the appropriate coatings for the best possible thermal isolation of the two zones.

Overall, THERMAP is thermally not very constraining for the spacecraft since it does not require cooling to a low temperature (<100 K) and it could be easily modified to be thermally regulated autonomously, with its own radiator if necessary (but with more mass as a drawback).

**Table 2 –** THERMAP main characteristics of the imaging and spectroscopic channels.

| Imaging channel | |
|---|---|
| Optical system | TMA |
| Field of view | 9.0° x 9.0° |
| Wavelength band pass | 8 – 18 µm |
| Focal length | 50 mm |
| Aperture | 25 mm |
| Focal ratio | 2 |
| Detector type | Microbolometers |
| Detector size | 640 x 480 pixels |
| Image size | 315 x 315 pixels |
| Pixel size | 25 µm |
| IFOV | 500 µrad |
| Spectroscopic channel | |
| Spectral dispersion system | Offner relay with a slit |
| Field of view (slit length x slit width) | 9° x 0.06° |
| Wavelength range | 8 – 16 µm |
| Spectral resolution (per pixel) | 0.3 µm ($\lambda/\Delta\lambda$ = 27 – 53) |
| Detector type | Microbolometers |
| Detector size | 640 x 480 pixels |
| Image size (slit length x spectral dispersion) | 315 x 40 pixels |
| Pixel size | 25 µm |
| IFOV (slit width) | 500 µrad |

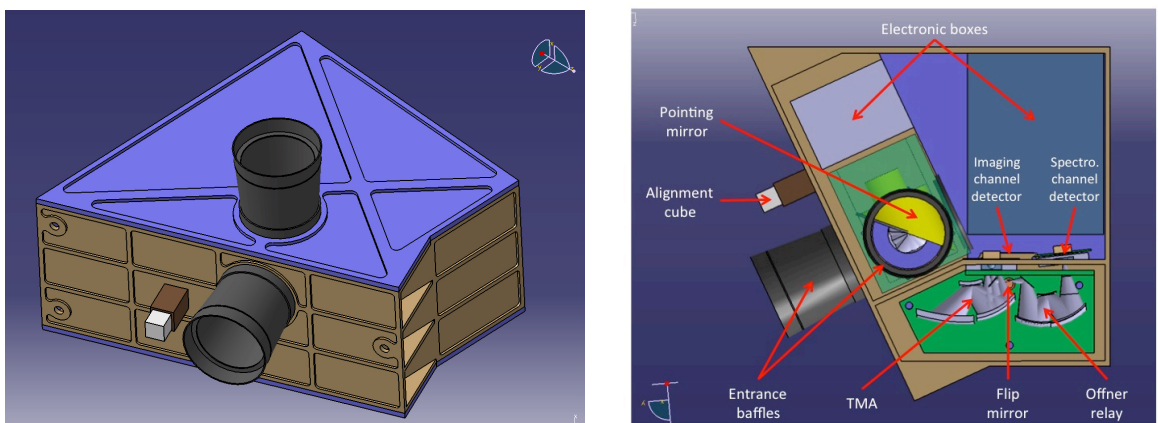

**Fig. 9 –** External (left panel) and internal view from above (right panel) of the optical and mechanical design of the THERMAP instrument. The optical design includes the pointing unit, which is a rotating mirror at the entrance of the instrument, followed by a TMA for imaging, followed by a flip mirror and a slit with an Offner relay for spectroscopy.

### 3.2.3 In flight instrument radiometric calibration

The calibration devices are of particular importance for accurate absolute radiometric measurements in the mid-infrared and constrain the instrument design. Our current solution is a rotating mirror, located at the entrance of the instrument, which will alternatively look at the asteroid, deep space, and two black bodies at a known temperature and located inside the instrument. Deep space observations are used to measure and then subtract the instrument thermal background, while black bodies are used as reference points. Calibration cycles are expected every few minutes, depending on the thermal stability of the instrument and therefore spacecraft attitude.

The first black body is passive. It is an aluminum plate with Acktar fractal black coating, equipped with temperature sensors PT100 or similar. The second black body is active; its temperature is adjustable. It has an emitting surface of 1.8 mm × 2.1 mm with a striped structure, made up of platinum on a special ceramic. The radiation source is an electrically heated resistor, consisting of a platinum structure whose resistance corresponds to a specific temperature. The temperature regulation procedure of the active black body uses a resistance bridge to keep the resistance of the radiation source constant. This procedure operates accurately under vacuum and keeps the surface of the active black body at a constant temperature.

### 3.2.4 Electrical design

The electrical design includes the following functional elements (Fig. 8):
- The Focal Plane Assembly (FPA), which contains the microbolometer and the minimum infrastructure required for its operation (PCB, proximity electronics, heat conductor and a flexible cable)
- The THERMAP Instrument Control Unit (ICU), which contains the electronics modules necessary for the management and operation of THERMAP (Data/telemetry, sequencing, configuration and modes, data acquisition and processing, housekeeping, temperature monitoring, control and monitoring):
    - Power Supply units, Main and Redundant (PS-M and PS-R)
    - Instrument Control Unit Processor Board (ICUP)
    - Front End Electronics (FEE)
    - HouseKeeping temperature monitoring (HK)
- Drivers Unit
    - Pointing Mirror Driver (PMD)
    - Flipping Mirror Driver (FMD)
    - Active Black Body Driver (ABBD)

### 3.2.5 Instrument resources

The resources of the THERMAP instrument are summarized in Table 3, including maturity margins plus 20% system margins.

**Table 3 –** Resources of the THERMAP instrument, including margins.

| Resources | |
|---|---|
| **Overall dimensions** | |
| Baffles MLI and connectors excluded | 28.7x25.5x14 cm |
| **Mass** | |
| Total | 7.5 kg |
| **Power** | |
| Observations | 20.4 W |
| Peak | 30.5 W |
| Standby | 4.2 W |
| **Data volume** | |
| Observations | 6.4 Gbit |
| Calibrations | 2.7 Gbit |
| Total | 9.1 Gbit |

| Temperature range | |
|---|---|
| Non operating | [ -40°C; +40°C] |
| Operating | [ +5°C; +15°C] |

## 3.3 Instrument performance

We performed a detailed analysis and numerous measurements to assess the performance of the THERMAP instrument. We provide here the main results for the imaging and spectroscopic channels.

### 3.3.1 Imaging channel

#### 3.3.1.1 Spatial resolution

The spatial resolution (2 pixels) of the imaging channel is defined by the IFOV (500 µrad):
- 10 m for far global characterization (from 10 km above the surface)
- 5 m for global characterization (from 5 km above the surface)
- 0.25 m for local characterization (from 250 m above the surface)

This spatial resolution is sufficient to meet scientific requirement #1 (Sect. 3.1).

#### 3.3.1.2 Optical performances

The image quality of the TMA telescope is excellent both in terms of spot diagram and modulation transfer function (MTF). The mean root mean square (RMS) spot radius is better than 1.25 pixels over more than 80% of the field of view and the MTF is optimized to yield high contrast images over the whole field of view. Efforts were made during the optical design of THERMAP to avoid high-order aspheric deformation terms on the mirrors, hence making their realization easier and their alignment less constraining.

#### 3.3.1.3 Temperature measurement accuracy

We measured the relative accuracy (NETD: Noise Equivalent Temperature Difference) of the detector for the THERMAP imaging channel, as illustrated in Fig. 10. The relative accuracy is better than 350 mK for temperatures above 250 K and can be estimated to better than 1 K for temperatures above 200 K.

We then performed an extensive analysis to estimate the error due to absolute calibration, depending on the number and temperature of the calibration points. Our conclusion is that THERMAP can measure absolute temperature with an accuracy better than 3.5 K above 200 K, assuming an absolute temperature uncertainty of 0.5 K on the internal calibration targets. The THERMAP imaging channel will thus provide accurate temperature measurements of the surface, sufficient to meet scientific requirement #1 (Sect. 3.1).

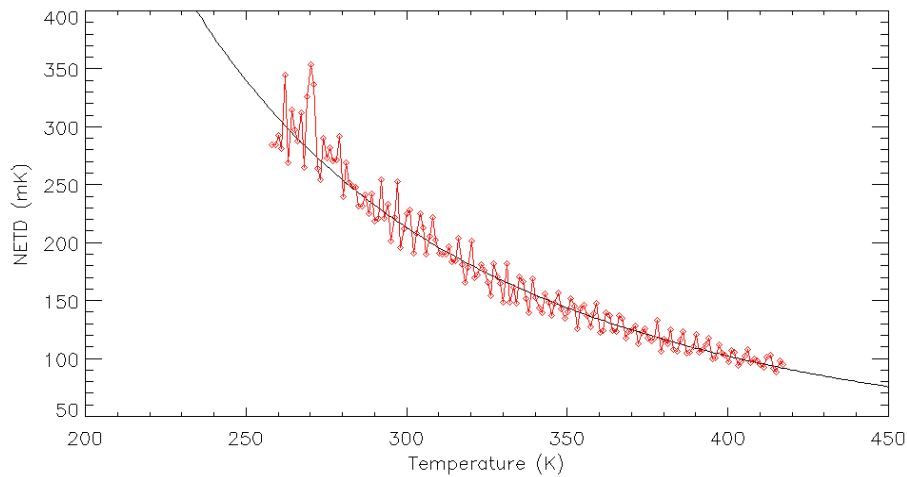

**Fig. 10** – Noise Equivalent Temperature difference (NETD) of the THERMAP detectors: 640x480 microbolometer array from the ULIS company. Measurements are in red and the fit in black (from Brageot et al. 2014).

### 3.3.2 Spectroscopic channel

#### 3.3.2.1 Spatial resolution

The spatial resolution of the spectroscopic channel is defined by the slit width (500 µrad), and is identical to that of the imaging channel since the Offner relay has a magnification of 1.0:
- 10 m for far global characterization
- 5 m for global characterization
- 0.25 m for local characterization

This spatial resolution is sufficient to meet scientific requirement #2 (Sect. 3.1).

#### 3.3.2.2 Optical performance

The image quality of the spectroscopic channel, including both the TMA telescope and the Offner spectrograph, is excellent both in terms of spot diagram and modulation transfer function (MTF). The mean root mean square (RMS) spot radius is smaller than one pixel over the whole field of view and for all wavelengths between 8 and 16 µm, and the MTF is optimized to yield perfectly resolved spectra. Once again, efforts were made during the optical design to avoid high-order aspheric deformation terms on the mirrors, hence making their realization easier and their alignment less constraining.

There is a small symmetric smile distortion of the spectra identical for all wavelengths, lower than 25 µm (1 pixel) from the centre of the slit to its edges. The data reduction pipeline can correct this distortion.

#### 3.3.2.3 Spectral resolution

An extensive analysis was performed by Hiesinger et al. (2010) in the framework of the MERTIS project (Bepi-Colombo ESA mission) to define the minimum spectral resolution required to perform the mineralogical studies of the surface of Mercury. They concluded that a spectral resolution better than 0.1 µm is required due to the complex mineralogy of Mercury. However, for a primitive asteroid, as in the case of MarcoPolo-R, the mineralogy is likely to be less complex than Mercury, and a spectral resolution of 0.1 µm is probably not required.

The THERMAP spectroscopic channel has a spectral resolution of 0.3 µm (per pixel) over the 8 – 16 µm wavelength range, i.e., $\lambda/\Delta\lambda$ = 27 – 53. We performed a detailed analysis, studying several types of minerals (Enstatite, Allende, Odessa, Ornans, Vigarano), and our conclusion is that this spectral resolution of 0.3 µm is indeed sufficient to reproduce the expected emission and absorption features of the main surface minerals (Fig. 11) and sufficient to meet scientific requirements #2 and #3 (Sect. 3.1).

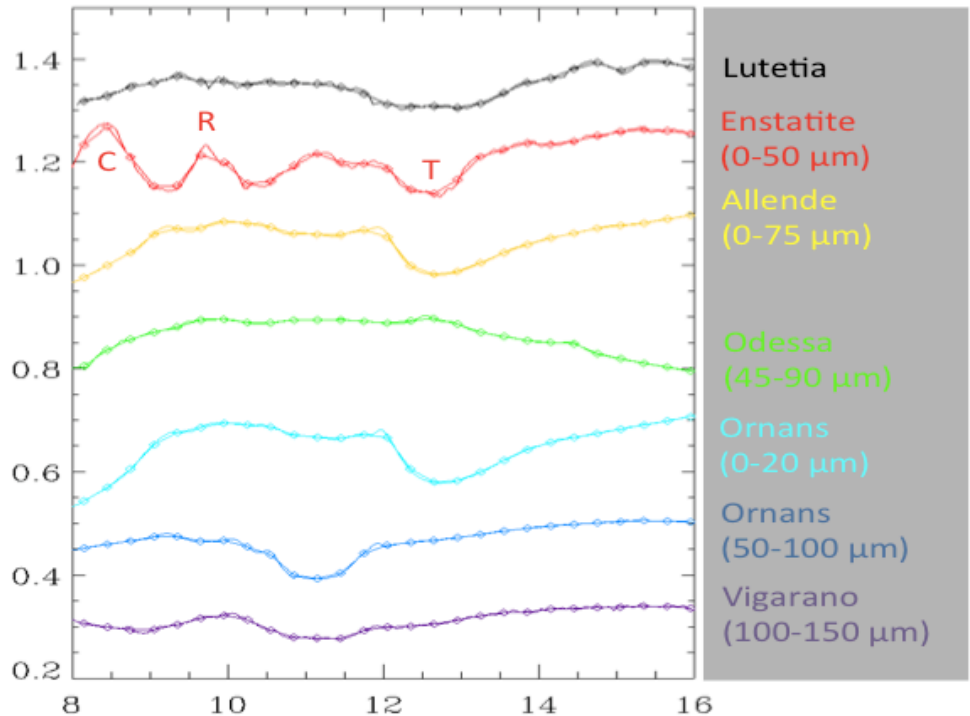

**Fig. 11** – Spectra of different minerals and meteorites. The symbols are spaced with the spectral resolution (0.3 µm) of THERMAP. Main features are easily detected, such as for example, Christiansen (C), Reststralhen (R) and Transparency (T) features. For each mineral and meteorite, the grain size range is specified (e.g., 0-50 µm for Enstatite).

### 3.3.2.4   Signal-to-noise of the spectra

We performed a study of the THERMAP spectroscopic performance using neutral density attenuation filters. The details of this analysis are available in Brageot et al. (2014). Following the MERTIS heritage, THERMAP will systematically acquire several spectra (typically 10 spectra) of a given region and average them on-board the spacecraft to increase the signal-to-noise-ratio. Only the averaged spectra will be sent to Earth. As illustrated in Fig. 12, the acquired spectra will have a SNR greater than 40 (limit for an exploitable spectra) between 9 and 13 µm for surface temperatures above 325 K, a SNR greater than 60 (limit for a good spectrum) between 9 and 13 µm for surface temperatures above 350 K, and a SNR up to 145 for surface temperatures above 400 K (as it is found for example around the sub-solar point at 1 AU from the Sun). This analysis shows that good spectra (SNR>60) of the asteroid surface can be acquired for all the regions located up to 50° from the sub-solar point. As the asteroid rotates, most of its surface will thus be observed in the spectroscopic channel. This is sufficient to meet scientific requirements #2 and #3 (Sect. 3.1).

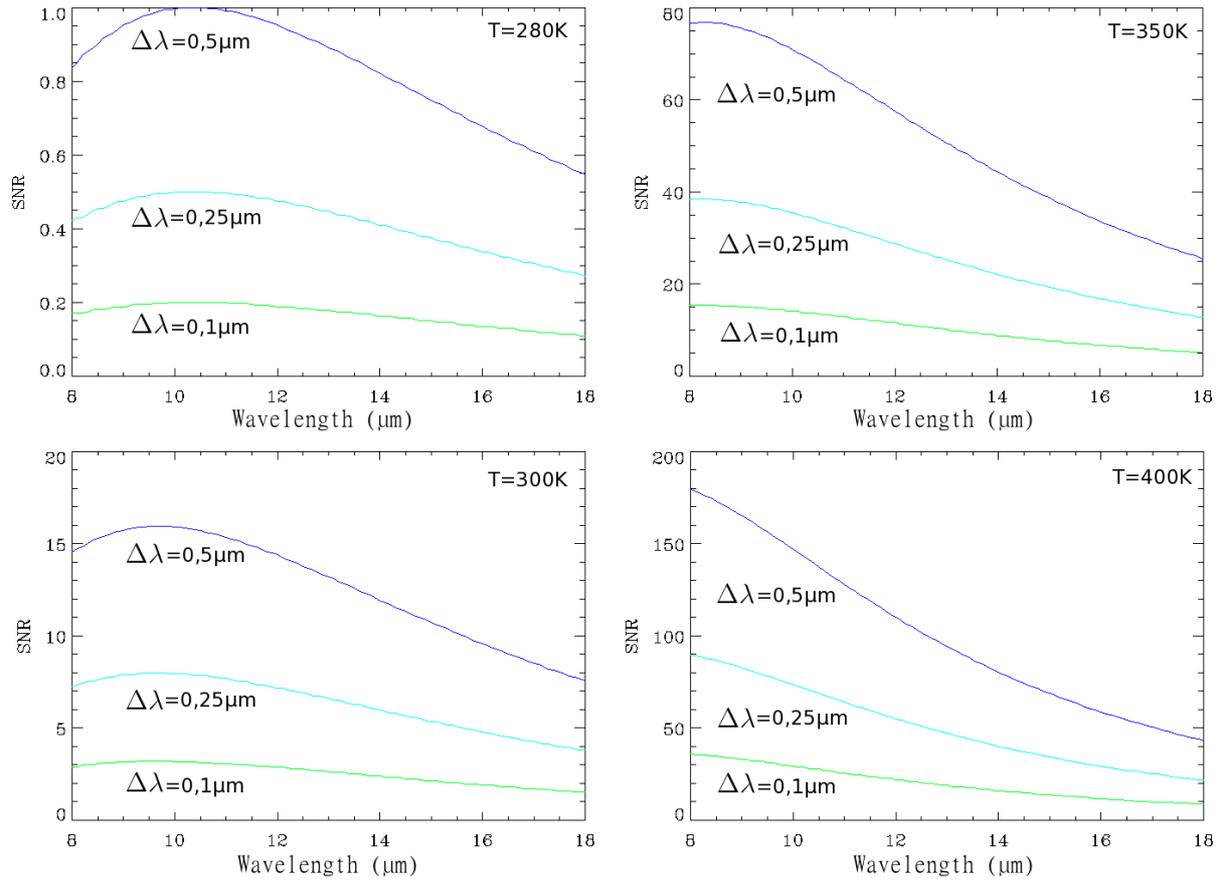

**Fig. 12** – Signal to noise ratio of the spectroscopic channel of the THERMAP instrument as a function of wavelength, for different spectral resolution and scene temperature (from Brageot et al. 2014).

### 3.4   Instrument heritage

THERMAP is a mid-infrared (8-16 µm) spectro-imager with two channels, one for imagery and one for spectroscopy. The THERMAP design follows the same philosophy as MERTIS, the Mercury Radiometer and Thermal Infrared Spectrometer for the ESA Bepi-Colombo mission (Hiesinger et al. 2010). It uses the same detector technology, i.e., uncooled micro-bolometer arrays, the same optical design, i.e., a three-mirror anastigmat telescope (TMA) for imaging followed by an Offner relay for spectroscopy, and the same principle for calibration, i.e., a rotating mirror at the entrance of the instrument that can point alternatively at the asteroid and three calibration targets (deep space and two black bodies).

The instrument electronic architecture is inherited from the work of the IAC on the development and realization of the electronics of the JEM-EUSO camera (JAXA). This camera uses the same uncooled micro-bolometer as THERMAP.

## 4   Conclusions and perspectives

We have presented THERMAP, a mid-infrared (8-16 µm) spectro-imager for space missions to small bodies in the inner solar system, developed in the framework of the MarcoPolo-R asteroid sample return mission. This instrument with two channels, one for imaging and one for spectroscopy, is very well suited for a NEO flyby or rendezvous. THERMAP has excellent imaging performance and reasonable spectroscopic capabilities. Since it is both a thermal camera with full 2D imaging capabilities and a slit spectrometer with a spectral resolution of 0.3 µm, THERMAP is a good instrument to characterize the surface thermal environment of a NEO and to map its surface composition.

The instrument is based on the new generation of uncooled microbolometer arrays. The relatively low spectral resolution of 0.3 μm of THERMAP could be improved using optics with smaller focal ratio (e.g., 1.4 instead of 2.0), replacing the reflective optics by refractive optics but keeping the same detectors. Another option could be to use cooled infrared detectors, but this would lose the benefits of the uncooled technology.

Although THERMAP will not fly on MarcoPolo-R since this mission was not selected by ESA, we hope to promote and propose the THERMAP instrument for future space missions towards airless bodies of the inner solar system.